\documentstyle[12pt]{article}
\newcommand{\al}{\alpha'}
\newcommand{\de}{\partial}
\newcommand{\be}{\begin{equation}}
\newcommand{\ba}{\begin{eqnarray}}
\newcommand{\ea}{\end{eqnarray}}
\newcommand{\ee}{\end{equation}}

\newcommand{\ca}{\mathcal}
\newcommand{\lr}{\leftrightarrow}
\newcommand{\f}{\frac}
\newcommand{\s}{\sqrt}

\newcommand{\ap}{\alpha}

\newcommand{\ddd}{\cdot\cdot\cdot}

\newcommand{\no}{\nonumber \\}
\newcommand{\la}{\langle}
\newcommand{\lb}{\rangle}
\newcommand{\ep}{\epsilon}

\newcommand{\beq}{\begin{equation}}
\newcommand{\eeq}{\end{equation}}
\newcommand{\beqa}{\begin{eqnarray}}
\newcommand{\eeqa}{\end{eqnarray}}
\newcommand{\CR}{\nonumber \\}

\begin{document}
\begin{titlepage}
\thispagestyle{empty}
\begin{flushright}
UT-02-13 \\
ITFA-2002-08 \\
hep-th/0203093 \\
March, 2002
\end{flushright}

\bigskip

\begin{center}
\noindent{\Large \textbf{
Strings on Orbifolded PP-waves}}\\
\vspace{2cm}
\noindent{
Tadashi Takayanagi\footnote{takayana@hep-th.phys.s.u-tokyo.ac.jp} }
and \, Seiji Terashima\footnote{sterashi@science.uva.nl}
\\
\vskip 2.5em

${}^1$
{\it Department of Physics, Faculty of Science, University of Tokyo\\
Hongo 7-3-1, Bunkyo-ku, Tokyo, 113-0033, Japan}

\vskip 2em

${}^2$
 {\it  Institute for Theoretical Physics,
University of Amsterdam \\
Valckenierstraat 65,
1018 XE, Amsterdam, The Netherlands}

\end{center}
\begin{abstract}

We show that
the string spectrum in the pp-wave limit of $AdS_5\times S^5/Z_M$
(orbifolded pp-wave) 
is reproduced from the ${\cal N}=2$ quiver gauge theory by
quantizing the Green-Schwarz string theory on the orbifolded
pp-wave in light cone gauge. We find that the twisted 
boundary condition on the world-sheet
is naturally interpreted from the viewpoint of
the quiver gauge theory.
The correction
of order $g_{YM}^2$ to the gauge theory operators
agrees with the result in its dual string theory.
We also discuss strings on some other orbifolded pp-waves.

\end{abstract}
\end{titlepage}

\newpage

\section{Introduction}

The AdS/CFT correspondence \cite{Ma, review}
gives us a deep understanding of duality between string theory and
gauge theory.
Furthermore, from this
we would know how the closed string theory
can be realized in terms of non-gravitational theory.
In the most interesting case $AdS_5 \times S^5$
the conjectured
duality to ${\ca{N}}=4$ super Yang-Mills theory has
been checked on the supergravity level \cite{GuKlPo,Wi}.
In order to show
this in full string theory we must analyze the world-sheet theory in the
presence of RR-flux. Recently, an intriguing
progress has been made in this direction \cite{BeMaNa}.
In the papers \cite{Me,MeTs} it was shown that the Penrose's limit \cite{Pe} 
of
$AdS_5 \times S^5$ in type IIB string theory 
is exactly solvable (see also related
papers \cite{RuTs,HaKaSa,BiPe}). The
background in this limit
is described by the maximally supersymmetric 
pp-wave (plane-fronted wave with parallel rays) 
with RR-flux  
\cite{BFHP2,BFP}.

The explicit metric of pp-wave \cite{BFHP} is given by
\ba
ds^2=-4dx^{+}dx^{-}-\mu^2
\sum_{i=1}^{8}(z^i)^2(dx^{+})^2+\sum_{i=1}^{8}(dz^i)^2,
\label{metric}
\ea
where we have defined the light-cone coordinate $x^{\pm}=\f{x^0\pm x^9}{2}$
and the RR-flux $F_{+1234}=F_{+5678}$ is proportional to
$\mu$. Using this solvable limit the authors of \cite{BeMaNa}
construct the explicit map between states of the string theory in pp-wave
background and the operators in the large $N$ limit of ${\ca{N}}=4$
super Yang-Mills theory which have
large  $U(1)$ R-charge $J\sim N^{\f12}$. 
These operators can be regarded
as `almost chiral primary' operators and the deviations from
chiral primary represent stringy excitations.

In this paper we would like to consider the pp-wave limit
of orbifolded $AdS_5 \times S^5$, especially 
$AdS_5 \times S^5/Z_M$
(see \cite{KaSi,LaNeVa,OzTe,Gu}
for the tests of duality on the supergravity level).
The orbifold action acts on the four coordinates $(z_6,z_7,z_8,z_9)$ in
the background (\ref{metric}).
This example is interesting because of two reasons. First we can discuss
the less supersymmetric ${\ca N}=2$ quiver gauge
theory\cite{DoMo,JoMy,KaSi,LaNeVa} (see also papers \cite{ItKlMu,GoOo,ZaSo}
for the enhancement of ${\ca N}=4$ supersymmetry in the pp-wave limit of 
${\ca N}=1$ gauge theory). Secondly an
orbifold is a good example of a stringy geometry and
thus it will be important to see how
its structure can be seen in the gauge theory side.
The duality in this orbifolded pp-wave was already implied in
\cite{BeMaNa} and the particular case $M=2$ was briefly
discussed in \cite{ItKlMu}.
Below we investigate the detailed correspondence for the $Z_M$
orbifolded string theory in
 both the untwisted sector and
twisted sectors. In particular we explicitly 
quantize the world-sheet theory on the
orbifolded pp-waves. This analysis reveals the interesting interpretation
of the twisted boundary condition in the orbifolded world-sheet theory
from the viewpoint of the ${\cal N}=2$ quiver gauge theory.
This enables us to
see that the string spectrum is exactly matched with the
gauge theory calculation up to order $g(\sim g_{YM}^2)$ correction.
We also mention a similar
analysis for more general orbifolds.

The contents of this paper is as follows. In section two we briefly 
review the results in \cite{BeMaNa}. In section three we quantize
the orbifolded string theory on pp-wave and discuss its correspondence
with the ${\ca N}=2$ quiver gauge theory. In section four we mention
general orbifolds. In section five we draw conclusions.

While preparing this paper for publication,
we received the preprints \cite{AlJa,KiPaReTh} which
have some overlap with ours.

\section{String on pp-waves and ${\cal N}=4$ super Yang-Mills }

In \cite{BeMaNa} it was proposed that the AdS/CFT correspondence \cite{Ma}
in the Penrose's limit is the duality between the string theory in the
pp-wave background (\ref{metric}) and the ${\ca N}=4$ super Yang-Mills
theory. The near horizon limit of a system of $N$ D3-branes is described by
the geometry $AdS_5\times S^5$ and its explicit metric is given by
\ba
ds^2=R^2\left(-dt^2\cosh^2 \rho+d\rho^2+\sinh ^2 \rho\ d\Omega^2_3
+d\psi^2\cos^2\theta+d\theta^2+\sin^2\theta d\Omega^{'2}_3\right),
\label{ads}
\ea
where $R=(4\pi gN\al^2)^{\f14}$ is the radius of $S^5$. 
For this background the Penrose's limit, which is generally defined as 
the limiting procedure of taking infinitely small 
neighborhoods of null geodesic, is given by the limit $R\to\infty$ 
focusing on $\rho=\theta=0$ with the following scaling\footnote{In taking 
this limit
one may consider that the periodicity of $\psi$ will lead to that of $x^+$.
One way to see that the periodicity is not important in the Penrose's limit
is to take another Penrose's limit discussed in \cite{RuTs}. We thank 
A.A.Tseytlin for pointing out this to us. 
We can show that even if we modify
the scaling (\ref{scl}) such that $x^+=\f{t+\lambda\psi}{1+\lambda}$ 
(others are not changed), we can obtain the same results (e.g.(\ref{metric})
and (\ref{delj})) 
in the 
region $\Delta\sim J$.}

\ba
x^+=\f{t+\psi}{2},\ x^-=\f{R^2(t-\psi)}{2},\ \rho=\f{r}{R},\ \theta=\f{y}{R}.
\label{scl}
\ea
Then we obtain exactly the pp-wave background (\ref{metric}) with $\mu=1$.

The pp-wave background (\ref{metric}) in 
light-cone Green-Schwarz(GS) string theory
is described by the following world-sheet Lagrangian \cite{Me,MeTs}
\ba
{\ca{L}}=\f{1}{2}\left[\de_{+}z^i\de_{-}z^i-(2\pi\al p^+\mu)^2
(z^i)^2\right],
\label{lag}
\ea
where $z^i\ \ (i=1,2,\ddd,8)$ are the scalar fields corresponding to
the spacetime coordinates not in the light-cone direction
and we omit the terms which include sixteen
fermions $S^{1 a}$ and $S^{2 a}\ \ (a=1,2,\ddd,8)$. 
In this paper we mainly show the analysis
of bosonic operators because the fermionic sector can be examined similarly.

An operator of conformal dimension 
$\Delta$ and R-charge $J$ in the gauge theory side corresponds to a state 
in the Green-Schwarz string which has the light-cone momentum $p^-,p^+$
by the following rule
\ba
2p^-=\Delta-J, \ \ 2p^+=\f{\Delta+J}{R^2}. \label{dual}
\ea
In order to
keep the value of $p^+$ and $p^-$ finite we focus on the operator
$\Delta\sim J\sim N^{\f12}$ in the large $N$ limit. We always
assume this scaling below in this paper.

In this duality \cite{BeMaNa} the ground state
$|\mbox{vac},p^-$=0$,p^+\lb$ of string theory
in pp-wave background
is identified with the operator $\mbox{tr}[Z^J]$
, which has the conformal dimension $\Delta=J$. The complex scalar field
$Z$ represents the transverse complex scalar field whose spin corresponds to
the value $J$.
The excitations in the string theory correspond to the insertions
of fields $\Phi^i$ (4 transverse scalars), $\chi_{J=1/2}^{a}$
(8 fermions)
 or covariant derivatives $D_{i}$ ($i=1,2,3,4$)
 in $\mbox{tr}[Z^J]$. The map between them is roughly given as follows
\ba
&&a_n^{\dagger i}\to D_{i}Z \ \ (i=1,2,3,4),\label{di}\\
&&a_n^{\dagger j}\to \Phi^{j} \ \ (j=5,6,7,8),\\
&&S_n^{\dagger a}\to \chi^a_{J=1/2} \ \ (a=1,2,\ddd,8),\label{fe}
\ea
where $a_n^{\dagger i}$ and $S_n^{\dagger a}$ denote
 the bosonic and fermionic oscillators with the level $n$
 (following the convention in \cite{BeMaNa})
 of the Green-Schwarz string in the
light-cone gauge.

The detailed map including the level $n$ is more complicated and shows the
way how the stringy excitations are translated into the gauge theoretic
counterparts \cite{BeMaNa}.
Let us consider the insertion of $\Phi^{i_j}$ operators $s$ times
into the $\mbox{tr} [Z^J]$ at just before
$l_j$-th $Z$ and summing over $l_j$
\beqa
&&\sum_{l_i=0}^{J-1} \prod_{i=1}^s
\exp \left( 2 \pi i \f{\l_i n_i}{J} \right)
{\rm tr}
\left( (Z)^{k_1} \Phi (Z)^{k_2} \Phi \cdots (Z)^{k_s} \Phi 
(Z)^{J-\sum_{i=1}^sk_s}
\right) \CR
&=& \sum_{l_i=0}^{J-1}
\exp \left( 2 \pi i \f{l_1 \sum_{j=1}^s  n_j}{J} \right)
\prod_{i=1}^s
\exp \left( 2 \pi i \f{(\l_i-l_1) n_i}{J}\right)
{\rm tr}
\left( (Z)^{k_1} \Phi \cdots (Z)^{k_s} \Phi (Z)^{J-\sum_{i=1}^sk_s}
\right) \no
\label{mop}
\eeqa
where the value $\sum_{j=1}^i k_j$ is equal to the $i$-th smallest value
among $l_j$'s.
Then it vanishes except $\sum_{i=1}^s n_i=0$
by summing up $l_1$ with fixing $l_i-l_1$
using the cyclic property of the trace. Thus we must require this relation.
Even though there is an ambiguity on the correct ordering at $l_i=l_j$,
it will not be important in the ``dilute gas'' approximation \cite{BeMaNa}.
Note that if one regards the positions of $\Phi^{i_j}$ as the physical 
positions of $s$ point particles in 1+1 dimensional spacetime $\Sigma$, 
the discrete Fourier transformation in (\ref{mop})
defines the `momentum' $n_i$ for each particle and the relation 
$\sum_{i=1}^s n_i=0$ is interpreted as momentum conservation. 
Furthermore, if we identify $\Sigma$ with the string would-sheet, then
the ``string of $Z$'s'' can be regarded as the physical string as
argued in \cite{BeMaNa}.

In string theory side we can interpret the operator (\ref{mop})
as
\ba
\prod_{j=1}^{s} (\ap^{\dagger i_j}_{n_j})|\mbox{vac},0,p^+\lb.
\ea
Note that the above state satisfies the level matching
condition $\sum_{i} n_i=0$ as expected. In string theory side we can
compute the light-cone energy
\ba
2p^-=\sum_{j=1}^s\s{1+\f{n_j^2}{(\al p^+)^2}}.
\ea

The duality tells us the relation (\ref{dual}) and we obtain
\ba
\Delta-J=\sum_{j=1}^s\s{1+\f{4\pi gNn_j^2}{J^2}}.\label{delj}
\ea
This prediction can be checked in the first order of the coupling constant
$g$ by the calculation in the large $N$ limit of 
${\ca N}=4$ super Yang-Mills theory \cite{BeMaNa}.

\section{Strings on $Z_M$ orbifolded pp-wave and
${\cal N}=2$ quiver gauge theory}

The low energy limit of $N$ D3-branes on 
the orbifold $C^2/Z_M$ can be described
by the ${\cal N}=2$ quiver gauge theory \cite{DoMo}, where the orbifold 
action
is defined as follows
\begin{eqnarray}
(\phi_1,\phi_2)\to (e^{\f{2\pi i}{M}}\phi_1,e^{\f{-2\pi i}{M}}\phi_2),
\label{ora}
\end{eqnarray}
where $\phi_1=z_6+iz_7$ and $\phi_2=z_8+iz_9$ denote
the coordinates of
$C^2/Z_M$ (see (\ref{metric})). In the near horizon limit we obtain 
the geometry $AdS_5 \times S^5/ Z_M$ and AdS/CFT correspondence tells us
the duality between this background and the ${\cal N}=2$ gauge theory 
\cite{KaSi,LaNeVa,OzTe,Gu}.
The metric of this background is given by the metric (\ref{ads}) 
orbifolded by the $Z_M$ action only on $\Omega'_3$
and the 
value of radius is given by $R=(4\pi gNM\al^2)^{\f14}$ because we start with
$MN$ D3-branes and further we impose the $Z_M$ projection. In this section
we would like to investigate the AdS/CFT correspondence by taking the
Penrose's limit.

\subsection{GS string on orbifolded pp-waves}

The Penrose's limit of $AdS_5 \times S^5/ Z_M$
is the $Z_M$ orbifold of the pp-wave background (\ref{metric}).
Since the orbifold projection does not act on the light-cone direction, 
we can describe GS string theory in this background simply by taking
the $Z_M$ orbifold in the sense of world-sheet theory in light-cone gauge
(refer to \cite{Me,MeTs} for the 
detailed analysis of GS string theory on pp-wave).
Then the orbifold action in string theory is defined by (\ref{ora}) 
identifying the coordinate $z^i$ and 
$\phi_i$ with the world-sheet scalar fields. 

The $m$-th twisted sector is defined by the
following boundary condition
\ba
&&z^i(\sigma+1,\tau)=z^i(\sigma,\tau), \ \ (i=1,2,3,4) \no
&&\phi_1(\sigma+1,\tau)=e^{\f{2\pi im}{M}}\phi_1(\sigma,\tau),
\ \ \ \phi_2(\sigma+1,\tau)=e^{-\f{2\pi im}{M}}\phi_2(\sigma,\tau).
\ea
We have also sixteen real fermions $S^{1a}$ and $S^{2a}$ and they
 are also divided into two groups.
The eight fermions in one group obey the trivial boundary condition
and the other eight are twisted by the phase $e^{\f{\pm 2\pi i}{M}}$.

For example, the mode expansions of (twisted) boson $\phi_1$ are given by
\ba
\phi_1(\sigma,\tau)=i\sum_{n\in {\bf Z}}(\f{1}{\omega _n}
e^{2\pi i(n-\delta)\sigma-i\omega _n\tau}\ap^{1L}_{n-\delta}
+\f{1}{\omega'_n}e^{-2\pi i(n+\delta)\sigma-i\omega' _n\tau}
\ap^{1R}_{n+\delta}),\no
\bar{\phi}_1(\sigma,\tau)=i\sum_{n\in {\bf Z}}(\f{1}{\omega' _n}
e^{2\pi i(n+\delta)\sigma-i\omega' _n\tau}\bar{\ap}^{1L}_{n+\delta}
+\f{1}{\omega_n}e^{-2\pi i(n-\delta)\sigma-i\omega _n\tau}
\bar{\ap}^{1R}_{n-\delta}),
\label{mode ex}
\ea
where $\delta=m/M$ and we can obtain the expression of $\phi_2$ by replacing
$\delta$ with $-\delta$.
Here we also defined
\ba
&&\omega _n=\pm 2\pi\s{(n-\delta)^2+(\al p^+ \mu)^2}\ \ \
\mbox{($+$ for $n>0$,
$-$ for $n\leq 0$)} \no
&& \omega' _n=\pm 2\pi\s{(n+\delta)^2+(\al p^+ \mu)^2}\ \ \
\mbox{($+$ for $n \geq 0$,$-$ for $n < 0$)}.
\ea

Then the usual canonical quantization of the Lagrangian (\ref{lag})
gives
\ba
&&[\ap^{1L}_{n-\delta},\bar{\ap}^{1L}_{n'+\delta}]=
\omega_n\delta_{n+n',0},\ \ \
[\ap^{1R}_{n+\delta},\bar{\ap}^{1R}_{n'-\delta}]=\omega_n'
\delta_{n+n',0}\ , \no
&&[\ap^{2L}_{n+\delta},\bar{\ap}^{2L}_{n'-\delta}]=
\omega_n'\delta_{n+n',0},\ \ \
[\ap^{2R}_{n-\delta},\bar{\ap}^{2R}_{n'+\delta}]=\omega_n
\delta_{n+n',0}.
\ea

The vacuum state of the world-sheet theory in the $m$-th twisted sector
is defined by
\ba
&&\ap^{1L}_{n-\delta}|\mbox{vac}\lb_m=
\bar{\ap}^{1R}_{n-\delta}|\mbox{vac}\lb_m=
\ap^{2R}_{n-\delta}|\mbox{vac}\lb_m=
\bar{\ap}^{2L}_{n-\delta}|\mbox{vac}\lb_m=0\ \
(n\geq 1),\no
&&\ap^{1R}_{n+\delta}|\mbox{vac}\lb_m=
\bar{\ap}^{1L}_{n+\delta}|\mbox{vac}\lb_m=
\ap^{2L}_{n+\delta}|\mbox{vac}\lb_m=
\bar{\ap}^{2R}_{n+\delta}|\mbox{vac}\lb_m=0\ \  (n\geq 0).
\ea

The bosonic part of
the light-cone Hamiltonian is given by

\ba
H&=&H_{0}+H_{\mbox{flat}}+\f{1}{2\pi\al p^+}\sum_{n=0}^{\infty}\left[
\ap^{1L}_{-n-\delta}\bar{\ap}^{1L}_{n+\delta}+
\bar{\ap}^{1R}_{-n-\delta}\ap^{1R}_{n+\delta}\right]\no
&&+\f{1}{2\pi\al p^+}\sum_{n=1}^{\infty}
\left[\bar{\ap}^{1L}_{-n+\delta}\ap^{1L}_{n-\delta}+
\ap^{1R}_{-n+\delta}\bar{\ap}^{1R}_{n-\delta}\right]
+(\phi_2\ \mbox{part}),
\ea
where $H_0$ and $H_{\mbox{flat}}$ denote
the contributions from zero-modes and four bosons in the
non-orbifold direction $(x^1,x^2,x^3,x^4)$. We omit these detailed forms
because both contributions are the same
as in \cite{MeTs}.

For example, the oscillators $\ap^{1L}_{-n-\delta}$ and
$\bar{\ap}^{1R}_{-n-\delta}$ in the $m$-th twisted sector have
the light-cone energy
\beq
2p^-=\sqrt{\mu^2+\frac{(n+\delta)^2}{(\al p^+)^2}}.\label{le}
\eeq
We assume that the zero-energy (or
the value of $p^-$ for the lowest state) in the twisted sector is zero and
will see that
the desirable spectrum is obtained in twisted sectors later. Note also 
that if we take the limit $\mu\to 0$, then we reproduce the mass spectrum of 
familiar string theory on the orbifold $C^2/Z_M$.

Then the spectrum should be $Z_M$ projected and also satisfy the level
matching condition. The level is defined to be the
summation of `$n\pm\delta$'
with respect to each oscillator for a given state. The level in the
left-moving
sector should be equal to that in the right-moving sector.

Now it is convenient to extract the creation operators
and we rename them as follows
\beqa
\ap^{\dagger 1}_{n} &\equiv &
\ap^{1L}_{n-\delta}  \;\; (n \leq 0),
\;\;\;\;\;\; \ap^{1R}_{-n+\delta}  \;\; (n >0), \CR
\bar{\ap^{\dagger 1}}_{n} &\equiv &
\bar{\ap}^{1L}_{n+\delta}  \;\; (n < 0) ,
\;\;\;\;\;\; \bar{\ap}^{1R}_{-n-\delta} \;\; (n \geq 0),\CR
\ap^{\dagger 2}_{n} &\equiv &
\ap^{2L}_{n+\delta}  \;\; (n< 0),
\;\;\;\;\;\; \ap^{2R}_{-n-\delta}  \;\; (n \geq 0), \CR
\bar{\ap^{\dagger 2}}_{n} &\equiv &
\bar{\ap}^{2L}_{n-\delta}  \;\; (n \leq 0) ,
\;\;\;\;\;\; \bar{\ap}^{2R}_{-n+\delta} \;\; (n> 0).
\label{rais}
\eeqa

Note also the $Z_M$ action
\ba
\ap^{\dagger 1}_{n}\to e^{2\pi i\delta}\ap^{\dagger 1}_{n},
\ \bar{\ap^{\dagger 1}}_{n}
\to e^{-2\pi i\delta}\bar{\ap^{\dagger 1}}_{n},\
\ap^{\dagger 2}_{n}\to
e^{-2\pi i\delta}\ap^{\dagger 2}_{n},\ \bar{\ap^{\dagger 2}}_{n}\to
e^{2\pi i\delta}\bar{\ap^{\dagger 2}}_{n},
\label{tran}
\ea
and the level of left-mover minus that of right-mover
for $\ap^{\dagger 1}_{n}, \bar{\ap^{\dagger 1}}_{n},
\ap^{\dagger 2}_{n}$ and $\bar{\ap^{\dagger 2}}_{n}$ are given by
$n-\delta, n+\delta, n+\delta$ and $n-\delta$, respectively.
Then we can construct all states
by operating these on the $m$-th twisted vacuum and
requiring the level matching condition
and invariance under the $Z_M$ action.
Note that there are two types of
$Z_M$ invariants, i.e.
a product of $M$ oscillators in (\ref{rais})
which have the same `$Z_M$ charge' of (\ref{tran})
and products of
 different `$Z_M$ charge' oscillators.
The differences of the levels of
these invariants are integer.

Finally let us briefly mention the fermionic zeromodes.
The orbifold projection
acts on the fermionic fields as they are space-time spinors.
In untwisted sector the fermion zero modes $S^{1a}_{0},S^{2b}_{0},
\ (a,b=1,2,\ddd,8)$
satisfy the
relation
$\{S^{ia}_{0},S^{jb}_{0}\}=\delta_{ab}\delta_{ij}$.
We can quantize these and get $2^{8}$ ``massless" states if we consider the
flat space not the orbifold.
In the orbifold theory the $Z_M$ acts on them as phase factor
$e^{2\pi i(s_3-s_4)/M}$
where $s_3$ and $s_4$ are spin along
$(x_6, x_7)$ and $(x_8, x_9)$, respectively,
and $s_3, s_4= \pm \frac{1}{2}$.
In the twisted sector we have eight fermion zero modes $S^{a}_{0}$ and
after the quantization we have $2^4=8 $bosons$+$8 fermions
``massless" states.
Four bosons parameterize the hyperkahler moduli of
the vanishing cycles and the other
 four bosons explain the RR field moduli if we set $\mu=0$.

\subsection{${\cal N}=2$ quiver gauge theory and string spectrum}

The quiver gauge theory
is defined by the corresponding quiver diagram \cite{DoMo,JoMy}.
In our case we consider
the quiver diagram of $A_{M-1}$ type, which has $M$ nodes and
arrows between them.
Each node represents a gauge group $U(N)$ and
we denote the corresponding $i$-th vector multiplet as
$(Z_i, W_i)$ in the ${\cal N}=1$ superfield notation.
Thus its gauge group\footnote{Note that in the context of AdS/CFT 
correspondence $N$ $U(1)$ factors of $G$ will be decoupled \cite{Gu}.}
 is $G=\prod_{i=1}^M U(N)$.
It should be also noted that the gauge coupling $g'$ 
of each $U(N)$ gauge theory
is given by $g'=Mg$ \cite{KaSi,LaNeVa}.

In the quiver diagram the arrow pointing one direction corresponds to
the superfield $Q^1_i, \bar{Q}^2_i$,
which is bifundamental matters $(\bar{N},N)$
with respect to $i$-th and $i+1$-th gauge groups,
and the other arrow to $Q^2_i, \bar{Q}^1_i$.
These form $M$ hyper multiplets $(Q_i^1, \bar{Q}_i^2)$,
$(i=1, \ldots ,M)$. Below we regard the index $i$ as $Z_{M}$
valued. The R-charges of these fields are summarized in Table 1.

\begin{table}[htbp]
\begin{center}
	\begin {tabular}{|c||c|c|c|}
\hline
  Field & $J=r/2$ & $2j$ & gauge charge \\
\hline
  $Z$ & 1 & 0 & adj \\
  $Q^1$ & 0& 1 & $(\bar{N},N)$ \\
  $Q^2$ & 0& 1 & $(N,\bar{N})$ \\
  $\bar{Q^1}$ & 0& -1& $(N,\bar{N})$ \\
  $\bar{Q^2}$ & 0& -1 & $(\bar{N},N)$ \\
\hline
\end{tabular}
\caption{R-charge}
\end{center}
\end{table}

Here $j$ and $r$ are the spin of $SU(2)_R$
and $U(1)_R$, respectively. This gauge theory has the
${\cal N}=2$ superconformal symmetry.
The chiral primary operators are operators with
$\Delta=2 j +r/2$ \cite{DoPe}. Below we also use the 
$Z_M$ projected $NM\times NM$
matrices $Z,Q^1,Q^2$ 
and they are defined as follows
\ba
Z=\left(
	\begin{array}{ccccc}
	Z_1 & 0 &\ddd & \ddd & 0  \\
	0 & Z_2 & \ddd  & \ddd & 0 \\
        \vdots & \vdots & \ddots & &  \vdots \\
         \vdots & \vdots & &  \ddots & \vdots \\
        0 & 0  & \ddd &\ddd &  Z_M 
	\end{array}
\right),\ \ 
Q=\left(
\begin{array}{ccccc}
	0 & Q_1 & 0 & \cdots & 0  \\
	0 & 0 & Q_2  & 0 & 0  \\
        \vdots & \vdots & \ddots & \ddots & \vdots \\
        \vdots  & \vdots & & \ddots &  Q_{M-1} \\
        Q_M & 0 & \cdots  & \cdots & 0
	\end{array}
\right).\label{zq}
\ea

After we take the pp-wave limit,
we have again the translation rule (\ref{dual})
and we should consider the operators which
have $\Delta \sim J \sim \sqrt{N}$ only.
{}From the Table 1, we see that they are
operators constructed from $J$ $Z$'s and
finite, but, arbitrary numbers of $Q$'s and
$\bar{Q}$'s.\footnote{
In this paper, we only consider the single trace operators.
The operators of $\Delta-J=2,3, \cdots,$ such as
$\bar{Z}, \partial Q, \cdots$
also are not considered
since we expect
such operators have a large conformal dimension
in $N \rightarrow \infty$ limit as argued in \cite{BeMaNa}.}
Then we can naturally identify
the translation rule of the gauge theory operators
into the operators in the string theory:
\ba
(\ap^{\dagger 1}_{n},\ \bar{\ap^{\dagger 1}}_{n},\
\ap^{\dagger 2}_{n},\ \bar{\ap^{\dagger2}}_{n})
\; \; \lr\ (Q^1,\bar{Q^1},Q^2,\bar{Q^2}).
\label{rule}
\ea
The translation rule with respect to
the other bosonic fields
in the gauge theory is
the same as in the ${\ca N}=4$ case (\ref{di}).

The light-cone energy
(\ref{le}) is rewritten in the gauge theoretic language as follows
\ba
\Delta-J=\s{1+\f{4\pi gNM(n+\delta)^2}{J^2}}=
\s{1+\f{4\pi g'N(n+\delta)^2}{J^2}},
\ea
Notice that t' Hooft coupling again appears correctly.

Next let us consider the correspondence between string states and
field theory operators in detail.
In the string theory, there are an untwisted
and $M-1$ twisted sectors.
To identify the corresponding operators we define the symbol $P$,
which acts on $Z, Q$ and $\bar{Q}$ such that
the $Z_M$ indices are shifted by 1.
{}From this,
we can define $M$ projectors
\beq
P_m \equiv \frac{1}{M} \sum_{j=0}^{M-1} \exp \left(
\frac{2\pi i m j}{M} \right) P^j.\label{proj}
\eeq

Using the projectors,
we can decompose any operator constructed from $Z, Q$ and $\bar{Q}$
as
\beq
{\cal O}_m=P_m {\cal O}(Z,Q,\bar{Q}).
\eeq
Then ${\cal O}_m$ is naively expected to correspond to the
$m$-th twisted sector.

The most simple example is the following operator
\ba
P_m {\mbox tr}\left[(Z_1)^J\right]
= \frac{1}{M}
\sum_{j=0}^{M-1}\exp\left( \frac{2\pi im j}{M} \right)
{\mbox tr}\left[(Z_j)^J\right],\label{zj}
\ea
which will couple to the $m$-th twisted sector
and correspond to the $m$-th twisted sector vacuum.
This speculation can be proved exactly as follows.
First note that the disk amplitude ${\cal A}$ of the m-th twisted field 
in closed string and the
open string massless field $Z$ is given\footnote{Here
we defined Tr as the trace for a $NM\times NM$ matrix, while
tr as the trace for a $N\times N$ matrix.} by
\ba
{\cal A} =\mbox{Tr}\left[\gamma_m\cdot Z^J\right]
\cdot \left\la V_{m}(V_{Z})^J\right\lb,\label{amp}
\ea
where $ V_{m}$ and $V_{Z}$ are vertex operators of the $m$-th twisted 
closed string and the
open string, respectively; $\la \ddd\lb$ denotes 
the correlation function and
the integration over moduli of world-sheet.
The Chan-Paton matrix $\gamma_m$ represents the
`Chan-Paton factor' for  $m$-th twisted sector as introduced in
\cite{DoMo,DiGo}. Its explicit form is the following diagonal matrix.
\ba
\gamma_m=\mbox{diag}
\left(1,e^{\frac{2\pi im}{M}},e^{\frac{4\pi im}{M}},\ddd,
e^{\frac{2(M-1)\pi im}{M}}\right).
\ea
Then the trace of Chan-Paton factor is given by
$e^{\frac{2\pi imj}{M}}$ and thus we finish the proof.

These simple operators (\ref{zj}) are chiral primary
 and their conformal dimensions are
given by $\Delta-J=0$. Note that for these states the spin of
R-symmetry $SU(2)_{R}$ vanishes. In the string theory side we can identify
${\cal O}_m$ with the oscillator vacuum state
$|\mbox{vac},p^-=0,p^+\lb_m$ in $m$-th twisted sector.

Next we would like to consider operators
corresponding to the excited states. We cannot insert a single
operator which has a non-zero $SU(2)_{R}$ charge because of
$Z_M$ projection. This is consistent with the string theoretic interpretation
(\ref{rule}). Note that this phenomenon is different from the ${\ca N}=4$ case
\cite{BeMaNa}, where we can insert such a single operator if the `momentum' 
vanishes $n=0$. Thus let us examine more than one excitation.
Then a naive guess is, for example,
simply following the procedures (\ref{mop}) and (\ref{proj})
\beq
\sum_{j=0}^{M-1}
\exp\left( \frac{2\pi im j}{M} \right) \times
\sum_{k=0}^{J}
\exp(\frac{2\pi i k n}{J})\times
{\rm tr} \left[(Z_j)^k Q' (Z_{j+1})^{J-k} Q'\right],
\label{mis}
\eeq
where $Q'$ represents $Q^i$ or $\bar{Q^i}$.
This is, however, incorrect
because they do not reproduce the level matching condition.
For instance, in $M=2$ case
the particular operator with $n=0$
\beq
\sum_{j=0}^{1}
\exp \left(\pi im j \right) \times
\sum_{k=0}^{J}
{\rm tr} \left[(Z_j)^k Q^1_j (Z_{j+1})^{J-k} Q^1_{j+1}\right].\label{van}
\eeq
vanishes except $m=0$
since the ${\rm tr [\cdots]}$ takes the same value for $j=0$ and $j=1$.
However, we have seen that there exists the
string state $\ap^{\dagger 1}_{0} 
\ap^{\dagger 1}_{1}|\mbox{vac},0,p^+\lb_{m=1}$, 
which satisfies
the level matching condition. 
Moreover, we can see that
the corrections to $\Delta-J$ of order $g$
are not reproduced by (\ref{mis}).
Thus this naive definition is problematic and
we should modify the identification.

Remembering the computation of previous 
disk amplitude (\ref{amp}) we would like to propose that the string state
\ba
\prod_{i=1}^s (\ap^{\dagger j_i}_{n_i})|\mbox{vac},0,p^+\lb_m, \label{ss}
\ea
corresponds to
the following 
operator instead of (\ref{mis}) 
\beqa
&&\sum_{l_i=0}^{J-1} \prod_{i=1}^s
\exp \left( 2 \pi i \f{\l_i (n_i+\epsilon_i \delta)}{J} \right)
{\rm Tr}
\left( \gamma_{m} Z^{k_1} Q' Z^{k_2} Q' \cdots Z^{k_s} Q' 
Z^{J-\sum_{i=1}^s k_i}
\right) \CR
&=& \sum_{l_i=0}^{J-1}
\exp \left( 2 \pi i \f{l_1 \sum_{j=1}^s  (n_j+\epsilon_j \delta)}{J} \right)
\prod_{i=1}^s
\exp \left( 2 \pi i \f{(\l_i-l_1) (n_i+\epsilon_i \delta)}{J} \right)\no
&& \times
{\rm Tr} 
\left( \gamma_{m} 
Z^{k_1} Q' Z^{k_2} Q' \cdots Z^{k_s} Q' Z^{J-\sum_{i=1}^s k_i}
\right), \label{tr}
\eeqa
where we use the same convention as in (\ref{mop}) and
$\epsilon_i$ is a sign of $Q'$'s fractional level\footnote{Here we
defined the values of $\epsilon_i$ such that $\epsilon_i=1$ for 
$\bar{Q}^1,Q^2$ and 
$\epsilon_i=-1$ for $Q^1,\bar{Q}^2$. Then $n_i+\delta\ep_i$
is equal to the level in (\ref{rais}).
}. The operators $Q'$ 
inserted in front of the $l_i$-th $Z$ are equivalent to the string 
state (\ref{ss})
via the rule (\ref{rule}). Note that here we used the trace Tr 
for $NM\times NM$ matrices and we assumed that 
the fields $Z$ and $Q'$ are $Z_{M}$ projected as in (\ref{zq}).
Then we can see that the level 
matching condition is
satisfied because the last line of (\ref{tr}) depends only on $(l_i-l_1)$.
Therefore we can find
the complete correspondence between the string spectrum and
operators in the gauge theory.

The shift of moding $n_i$ by $\epsilon_i \delta$ in the above operator
can be explained as follows. It should be required that the operator
like (\ref{tr}) is invariant under the shift of the position $Q'$. In
particular we can shift the value of $l_i$ by $J$ which corresponds to
the rotation of $Q'$ in the trace.
This is because
we can interpret $l_i/J$ as the world-sheet periodic coordinate
$\sigma$ in (\ref{mode ex}) following the general idea of \cite{BeMaNa} that
the ``string of $Z$'s'' can be regarded as the physical string. 
If we move one of the fields $Q^1,\bar{Q^2}$ (or $Q^2,\bar{Q^1}$) 
so that it crosses the matrix $\gamma_{m}$, then we have the extra factor
$e^{2\pi i\delta}$ (or $e^{-2\pi i\delta}$). A typical example is shown below
\ba
{\rm Tr}\left[\gamma_m Z^k\bar{Q}^1Z^{J-k}Q^1\right]
=e^{2\pi i\delta}\ {\rm Tr}\left[\gamma_m Z^{J-k}Q^1 Z^k\bar{Q}^1
\right],
\ea
where we have employed the commutation relation 
$Q^1\gamma_m=e^{2\pi i\delta}\gamma_mQ^1$.
In order to cancel this extra
factor we must shift the value of $n_i$ by $\epsilon_i \delta$. 
Actually under the shift of $l_i$ the summation of (\ref{tr}) is invariant.
In other words, the presence of the matrix $\gamma_m$ is equivalent to
the existence of Wilson line in the `periodic direction $\sigma$'
and thus 
it shifts the value of momentum. In this way the quiver gauge theory 
operator (\ref{tr}) 
explicitly shows the twisted boundary condition in the string theory of 
the orbifolded  pp-wave!

Let us examine the operator (\ref{tr}) in the
explicit examples which
include two hyper multiplet fields.
There are four possibilities
of hyper multiplets: $Q^1\bar{Q^1},\ Q^2\bar{Q^2},\ Q^1Q^2$ and
$\bar{Q^1}\bar{Q^2}$. These states each
correspond to the following four states in pp-wave string theory
(see the rule (\ref{rule}))
\ba
\ap^{1}_{n}\bar{\ap}^{1}_{-n}|\mbox{vac}\lb,\ \
\ap^{2}_{n}\bar{\ap}^{2}_{-n}|\mbox{vac}\lb,\ \
\ap^{1}_{n}\ap^{2}_{-n}|\mbox{vac}\lb,\ \
\bar{\ap}^{2}_{n}\bar{\ap}^{1}_{-n}|\mbox{vac}\lb.
\ea
Note that these states satisfy the level matching condition and
the $Z_M$ invariance.

We can construct the dual operator in the gauge theory side by using
the general formula (\ref{tr}). Then we obtain the following result
in the $Q^1Q^2$ case (up to the normalization factor) after taking 
the trace with respect to $j=0,1,\ddd,M-1$
\beq
\sum_{j=0}^{M-1}
\exp\left(2\pi ij\delta \right) \times
\sum_{k=0}^{J-1}
\exp\left(\frac{2\pi ik \left( n- \delta\right)}{J}\right)\times
\mbox{tr}\left[(Z_j)^k Q^1_j (Z_{j+1})^{J-k} Q^2_j \right],
\eeq
and similar results can be shown in the other three cases.

We can check that this operator has the correct conformal dimension
in the large $N$ limit
\ba
\Delta-J=2\sqrt{1+\frac{4\pi g'N(n-\delta)^2}{J^2}}
\simeq 2+\frac{4\pi g'N(n-\delta)^2}{J^2},
\ea
to the first order in the coupling $g'\sim g_{YM}^2$ as in the paper
\cite{BeMaNa}\footnote{There is a subtle point in this argument
for the operators which are not chiral.
Here we assume the momentum independent 
contributions vanish if $n=m=0$.}. 
The correction
proportional to
$g'N/J^2$ comes from the interaction 
$\sim \sum_{j=0}^{M-1}(Z_jQ^1_j\bar{Z}_{j+1}\bar{Q}^1_{j+1}
+Z_{j+1}\bar{Q}^1_{j+1}\bar{Z}_jQ^1_j)+(Q^2Z\bar{Q}^2\bar{Z}\ \mbox{term})$.
We have also used the ``dilute gas'' 
approximation as in \cite{BeMaNa}\footnote{One can see that the 
interaction of the form 
$Q\bar{Q}Q\bar{Q}$ does not affect our arguments in the dilute gas 
approximation. Then we expect this term can be ignored.} .

We can also generally check that the spectrum of the string theory
is consistent with the quiver gauge theory result
at least up to $g'$ order for the operator form (\ref{tr})
\ba
\Delta-J=\sum_{i=1}^{s}\sqrt{1+\frac{4\pi g'N(n_i+\ep_i\delta)^2}{J^2}}
\simeq s+\sum_{i=1}^{s}\frac{2\pi g'N(n_i+\ep_i\delta)^2}{J^2}.
\ea

Finally we would like to mention 
the gauge theory dual of the string states
which include the fermionic operators 
$S^{\dagger a}_n\ \ (a=1,2,\ddd,8)$. 
There are four fermionic operators which are $Z_M$ invariant. They 
correspond to fermions in the vector multiplets of ${\ca N}=2$ 
quiver gauge theory. The other four operators which are 
not $Z_M$ invariant correspond to fermions in the hyper multiplets.
Both of them can be treated in the same way as $Z$ and $Q'$, respectively and
thus we omit the detailed discussion.

\section{General orbifolds}

Here let us briefly discuss the generalization of previous results
for other two dimensional orbifolds $C^2/\Gamma$.
First consider non-abelian supersymmetric orbifolds which are classified
by $A$(=$Z_M$),$D$,$E$ series \cite{JoMy,DiGo}.
In these
orbifolds each twisted sector is labeled by the conjugacy class
${\ca C}_\beta,\ \ (\beta=1,2,\ddd,r)$ of $\Gamma$. The quiver gauge
theories
correspond to the regular representation $\rho_{reg}=\oplus_{\ap=1}^{r}
n_{\ap}\rho_{\ap}$,
where $\rho_{\ap} \ \
(1\leq\ap\leq r)$ are irreducible representations of $\Gamma$ and
we defined $n_{\ap}\equiv\mbox{dim}\rho_{\ap}$.
Their gauge groups are given by $G=\prod_{\ap=1}^{r}U(Nn_\ap)$ and each
gauge coupling
is given by $\tau_{\ap}=n_\ap\tau/|\Gamma|\ \  (\tau=i/g+\chi)$.
In order to know the coupling of fields in the gauge theory
to the ${\ca C}_\beta$ twisted sector we have only to replace
(\ref{proj}) with
\ba
\frac{1}{|\Gamma|} \sum_{\ap=1}^{r} \chi_{\ap}({\ca C}_\beta)
P_\ap,
\ea
where $\chi_{\ap}(g)=\mbox{tr}_{\ap}(g)$ denotes the character of an element
of
$g\in  \Gamma$ in $\rho_{\ap}$ representation. The symbol
$P_\ap$ denotes the projection onto the gauge group $U(Nn_\ap)$.
All other arguments (e.g.(\ref{tr})) in the 
section three can also be generalized 
for $D$ and $E$ cases.

It is also interesting to consider non-supersymmetric orbifolds.
For example, let us
consider the $Z_M$ orbifold defined by
\ba
\phi_1\to e^{\f{2\pi i}{M}}\phi_1,\ \ \ \phi_2\to e^{-\f{2\pi iL}{M}}\phi_2,
\ea
where $L$ is an odd integer and except for the cases $L=\pm 1$
the supersymmetry is completely broken. This orbifold was recently discussed
in the context of closed string
tachyon condensation \cite{AdPoSi}.
Then
we can speculate the same identification (\ref{rule}) in the untwisted sector.
Here we should regard
scalar fields $Q^1$ and $Q^2$ as $(Q^1)_{j,j+1},(Q^2)_{j+L,j}$
instead of $(Q_1)_{j,j+1}(=Q^1_j),(Q_2)_{j+1,j}(=Q^2_j)$, where we have
written the indices of Chan-Paton matrices explicitly.
The quantization of GS string on this nonsupersymmetric background
is also the same as in section 3.1.
In twisted sectors, however, the spectrum
of string is slightly difficult to analyze because of the ambiguity of zero
energy.
In any case it will be an intriguing fact that we can find string states 
in untwisted sector
whose light-cone energy is independent of the coupling $g$.
Thus there seem to exist the operators
whose conformal dimensions are fixed as $\Delta-J=0,1,2,\ddd$ in the large $N$
limit of the non-supersymmetric gauge theory
if we believe the
AdS/CFT correspondence in such a pp-wave limit.

\section{Conclusions}

In this paper we investigated an extension of the
AdS/CFT correspondence in the limit
of the orbifolded pp-wave background. We examined both the spectrum of 
states in the orbifolded 
Green-Schwarz string and the operators in the gauge theory side.
The results
show that the duality
between string theory and ${\ca N}=4$ 
super Yang-Mills theory discussed in \cite{BeMaNa}
holds for less supersymmetric case (${\ca N}=2$ quiver gauge theory). 
We found an interesting 
interpretation of the twisted boundary condition in the orbifold theory
from the viewpoint of the quiver gauge theory. We also discussed 
the similar issue in more general orbifolds and found that some parts of 
the results can be applied to non-supersymmetric orbifolds. It will be
also interesting to analyze the pp-wave limit of D3-branes on
the orbifolds $C^3/\Gamma$
, which will lead to ${\cal N}=1$ quiver gauge theory (for some previous 
discussions see \cite{GoOo}).

\vskip6mm
\noindent
{\bf Acknowledgments}

\vskip2mm
We thank A.A.Tseylin for useful comments.
The research of T.T
was supported in part by JSPS Research Fellowships for Young
Scientists.

\end{document}